\documentclass[a4paper]{article}
\usepackage[a4paper,top=3cm,bottom=2cm,left=3cm,right=3cm,marginparwidth=1.75cm]{geometry}
\usepackage[utf8]{inputenc}

\usepackage{natbib}
\setcitestyle{authoryear,open={(},close={)}}

\usepackage{amsmath}
\usepackage{graphicx}
\usepackage[colorinlistoftodos]{todonotes}
\usepackage[colorlinks=true, allcolors=blue]{hyperref}

\def\R{\mathbb R}

\def\bSig\mathbf{\Sigma}

\newcommand{\1}{\mathbf{1}}
\newtheorem{defn}{{\bf Definition}}
\newtheorem{thm}{{\bf Theorem}}
\newtheorem{cor}{{\bf Corollary}}
\newtheorem{pro}{{\bf Proposition}}
\newtheorem{lmm}{{\bf Lemma}}

\newcommand{\EQ}[1]{\begin{eqnarray}#1\end{eqnarray}}
\renewcommand{\theequation}{\arabic{section}.\arabic{equation}}

\usepackage{amsmath,graphicx,amssymb,braket,xcolor,subfigure,upgreek,bbold}
\usepackage{calrsfs}

\DeclareMathAlphabet{\pazocal}{OMS}{zplm}{m}{n}
\newcommand{\La}{\mathcal{I}}

\usepackage[hints,insection]{minitoc}
\mtcsetdepth{secttoc}{2}

\begin{document}

\title{The nph2ph-transform: applications to the statistical analysis of completed clinical trials}
\author{Sean M. Devlin$^1$ and John O’Quigley$^2$}
\date{%
    $^1$Memorial Sloan Kettering Cancer Center, New York, USA\\%
   $^2$Department of Statistical Science, University College London, U.K. 
    \today
}


\maketitle

\begin{abstract}
{We present several illustrations from completed clinical trials on a statistical approach that allows us to gain useful insights regarding the time dependency of treatment effects. Our approach leans on a simple proposition: all non-proportional hazards (NPH) models are equivalent to a proportional hazards model. The nph2ph transform brings an NPH model into a PH form. We often find very simple approximations for this transform, enabling us to analyze complex NPH observations as though they had arisen under proportional hazards. Many techniques become available to us, and we use these to understand treatment effects better.}

{Brownian motion, clinical trials,  model fit, proportional hazards, time-dependent effects}
\end{abstract}

\maketitle

\section{Introduction}

There are two main goals to a clinical research study that involves survival as a main endpoint. The first goal, an important one although relatively easy to express, is to decide if some experimental treatment results in prolonging survival. The log-rank test plays a key role in such studies and typically forms the basis for sample size calculations given a Type 1 error and hoped-for power under an alternative hypothesis that the experimental treatment does have an impact. To carry out such calculations, some simplifying assumptions need to be made, the most common being that if we can reject the null hypothesis, the alternative will be well approximated by a proportional hazards model. The great majority of studies make such an assumption and will sometimes include in the protocol some formal way of testing the validity of the assumption. 

The second main goal will be addressed once the trial and the main results are completed. The subsequent analysis is no longer restricted by the design structure, the hypotheses of interest, and the desire to maintain careful control on statistical error. We are at liberty to carry out a broader investigation of the treatment effect. While conclusions from such secondary investigations will lack the impact concerning the main objectives and endpoints, they can still be of considerable value in model building. The purpose of model building may be to gain greater insight and understanding of the myriad of mechanisms that can affect survival or to obtain some pointers into ways in which we might refine and improve our initial design. This can help plan future related studies.  

Our focus here is on this second goal. There is a rich literature on prognostic studies and model building \citep{Harrell1996,  Gerds2008,Steyerberg2013} and this can be an important component of the retrospective analysis of a clinical trial. We do not consider this here since our focus is somewhat narrower. We consider the study itself, most often involving two comparative groups, and investigate the impact of the proportional hazards assumption and to what extent predictive power can be increased when survival is more accurately modeled by a non-proportional hazards assumption.  

\begin{figure}[h]
\centering
 \includegraphics[width=3in, height=5in]{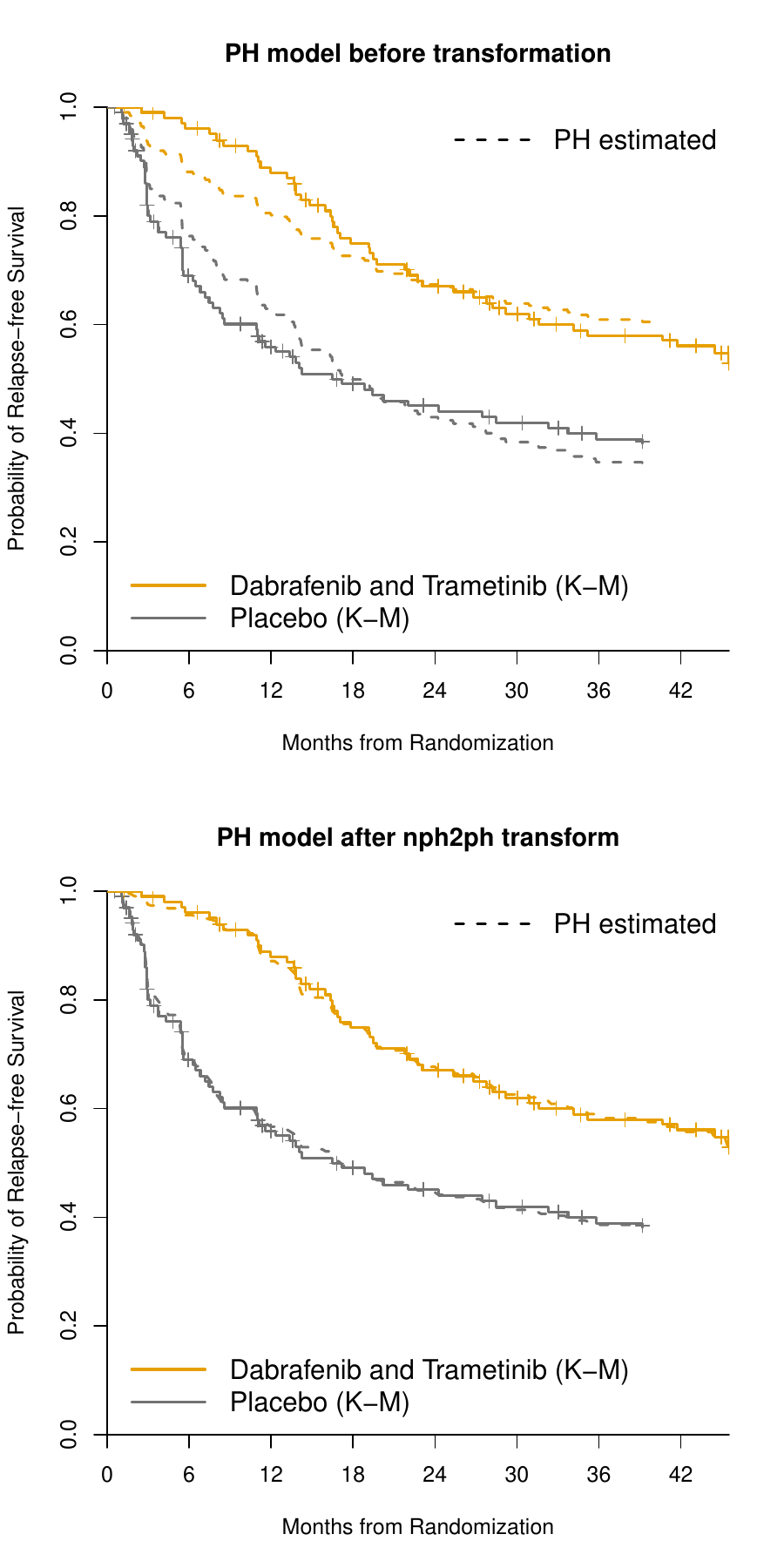}
 \caption{Kaplan-Meier survival curves from a randomized trial in melanoma \citep{LongNEJM2017}. The figure on the top shows the estimated conditional survival functions obtained under the best-fitting proportional hazards model. The bottom figure shows the estimated conditional survival functions under the best-fitting piecewise proportional hazards model after implementing a nph2ph transform.   \label{fig.LongIntro}   }
\end{figure} 
 
 A recent study (Figure \ref{fig.LongIntro}) published in the New England Journal of Medicine showed the clear advantage of dabrafenib combined with trametinib over placebo in melanoma patients with a BRAF mutation. The outcome was relapse-free survival, and, in this regard, the superiority of the treatment was clear. An assumption of PH looks to be a very reasonable one. However, the PH model is not a good fit, as shown later in Section \ref{sec.trialdab}. The simplest possible approximation, as discussed in Section \ref{sec.approxp}, to the unknown nph2ph transform - a piecewise PH model with a single changepoint - dramatically improves fit. This is immediately visible in Figure \ref{fig.LongIntro} where not only does the transformed model provide a much more accurate summary of the Kaplan-Meier curves, but we also note that there is minimal room for further improvement.   

The strength of the treatment effect, measured under an incorrect PH assumption, will underestimate the true strength of the effect when treatment effects are time-dependent  \citep{Chauvel2017}. The bias can be large, and we will see many situations in which the actual treatment effect can be double, or even triple, what we had thought it to be due to a poor model specification. This is a very important finding in its own right and enables us to provide a more relevant summary of the trial's outcome. The fact that a trial's design may have led to the underestimation of treatment effect, along with an indication of how much, can be a valuable consideration in deciding how best to move forward. We mention in passing that the power of many statistical tests, the log-rank test, in particular, depends on the PH assumption. Under a more accurate model specification, we may have achieved greater power.   
 
The nph2ph transform is very much a data analytic tool. By this, we mean that we put less emphasis on theoretical population models and the resulting distribution theory and more on the data. Our starting point is the observed data rather than some postulated model. We introduce an index of fit, which measures the degree to which observed data can be taken to have a PH form. Our interest is not in testing the fit - although this could be done - but in deciding how close to or far from a PH structure the data are. And, in using the nph2ph transformation, we can make NPH observations appear more like PH observations.

\section{Basic methodology}

\subsection{Proportional and non-proportional hazards models}\label{sec.struc}

The random variables of interest are the failure time, $T_i$, the
censoring time, $C_i,$ and the treatment indicator covariate,
$Z_i(\cdot)$, $i=1,...,n.$ 
Generally, the treatment indicator, $Z_i (t),$ will remain constant in time. Allowing for time dependency creates no additional methodological difficulties; in certain situations, for example, cross-over trials, this may be used to advantage. 
We use $\La ( A )$ as an indicator function so that, $\La (A)=1$ when $A$ is true and is zero otherwise.
Let $F(t)=P(T<t)$.
For each subject $i$ we observe  $X_i=\min(T_i, C_i)$, and
$\delta_i= \La (T_i\leq C_i)$. The ``at risk'' indicator $Y_i(t)$ is
defined as, $Y_i(t)= \La (X_i\geq t).$ The counting process $N_i(t)$
is defined as, $N_i(t)=\1 \{T_i\leq t, T_i\leq C_i\}.$ $\bar{N}(t)=\sum_1^n N_i(t)$ and we transform the failure times to the interval (0,1) via the homomorphism $\phi (t),$ described below.   
Closely related to $Z$ is the observation at $t$, ${\cal Z }_t= \sum_{i=1}^n Z_i\left(X_i\right) \La \{\displaystyle \phi_n(X_i)=t,\delta_i=1 \},\quad t\in[0,1], 
$ in words, an outcome that assumes the value of the covariate (group assignment) associated with the failure, when there is one, at time $t,$; otherwise, it assumes the value zero.  \cite{jmoq03} and \cite{Chauvel2014} chose to suppress the random nature of $T,$ treating it as an indexing variable $(1/k_n, 2/k_n, ... )$ on the interval (0,1) so that we can now view ${\cal Z }_t$ as a stochastic process, i.e., a collection of random variables indexed by $t.$  To complete the characterization of the process, we associate with each $t,$ the distribution $\pi_i(\beta(t),t).$ While any well-defined distribution could be used, we focus on the specific definition  
\begin{eqnarray}
\pi_i(\beta(t), t) = \frac{Y_i(t) \exp\{\beta(t) Z_i(t) \} }{ \sum_{j=1}^n Y_j(t) \exp\{\beta(t) Z_j(t)\} }, \text{ } i=1,\hdots, n,  
\end{eqnarray}
which arises naturally in the setting of proportional and non-proportional hazards models. 

We group the models under as general a heading as possible. The most general model is then the non
proportional hazards model written as
\begin{equation}
\label{nonph}
\lambda(t|Z(t))=\lambda_0(t)\exp\{\beta(t) Z(t)\}, 
\end{equation}
 where
$\lambda(t|\cdot)$ is the conditional intensity function,
$\lambda_0(t)$ the baseline hazard and $\beta(t)$ the time-varying
regression effect. 
The above model becomes a proportional hazards model under the
restriction that $\beta(t)=\beta$, a constant i.e. 
$\lambda(t|Z(t))=\lambda_0(t)\exp\{\beta Z(t)\}. $  
In the proportional hazards model, a strictly monotonically increasing transformation in time does not affect inference of the regression coefficient $\beta$. Indeed, the time of death and censored values are not used for inference; only their order of occurrence counts. The transformation of \cite{Chauvel2014}  will also conserve the order of observed times while simultaneously leading to great tractability. By applying the inverse transformation, we can go back to the initial time scale. More precisely, the transformation of \cite{Chauvel2014} can be seen as an isomorphism $\phi_n$ on the vector $X_1,\dots,X_n ,$ taking us from the observed time outcomes to the interval (0,1). The observed failures map to the values $(1/k_n, 2/k_n, ...,)$ where $k_n$ equates to the observed number of failures in the simplest case. More generally, we will ignore failures that contain no information relating to treatment effect, i.e., the set of the larger failures taking place when the risk set only contains individuals from the same treatment group.

\subsection{Treatment effect process}
A key ingredient in building the nph2ph transform is the treatment effect process, which is no more than a special case of the regression effect process \citep{Chauvel2014, Chauvel2017, Flandre2019, jmoq21} applied to the situation of two treatment groups. Expectations and variances with respect to the family of probabilities $\{\pi_i(\beta(t),t),i=1,\dots,n\}$ can now be defined, and these are all that are needed to define the treatment effect process. 
\begin{defn}
The moments of $Z$ with respect to the family of probabilities $\{\pi_i(\beta(t),t),i=1,\dots,n\}$ and in which $t \in[0,1] ,$ are given by,
\begin{equation} \label{esp_stand} 
{\cal E }_{\beta(t)}(Z^m \mid t) = \sum_{i=1}^n Z^m_i\left(\phi_n^{-1}(t)\right) \pi_i(\beta(t),t),
\end{equation}
providing us with the mean ${\cal E }_{\beta(t)}(Z \mid t)$ and variance, ${\cal V }_{\beta(t)}(Z\mid t) = {\cal E }_{\beta(t)}(Z^2 \mid t) - {\cal E }^2_{\beta(t)}(Z \mid t) .$
\end{defn}
We use this mean and variance to construct the following standardized process that we will refer to as the treatment effect process. 
\begin{defn}
Let $j=0,1,\dots,k_n$ and $t_j = j/k_n .$ Given $\beta(t_j),$ the treatment effect process at time $t_j$ is written: 
\begin{equation}
\label{brown1u}
U^*_n\left( \beta(t_j), t_j\right) =
\dfrac{1}{\sqrt{k_n}}\sum_{i=1}^{j} {\cal V}_{\beta(t_i)} (Z\mid t_i)^{-1/2} \left\{\mathcal{Z}(t_i)-{\cal E}_{\beta(t_i)} (Z\mid t_i)\right\}.
\end{equation}
\end{defn}

\begin{figure}[h]
\centering
 \includegraphics[width=5.0in, height=5.0in]{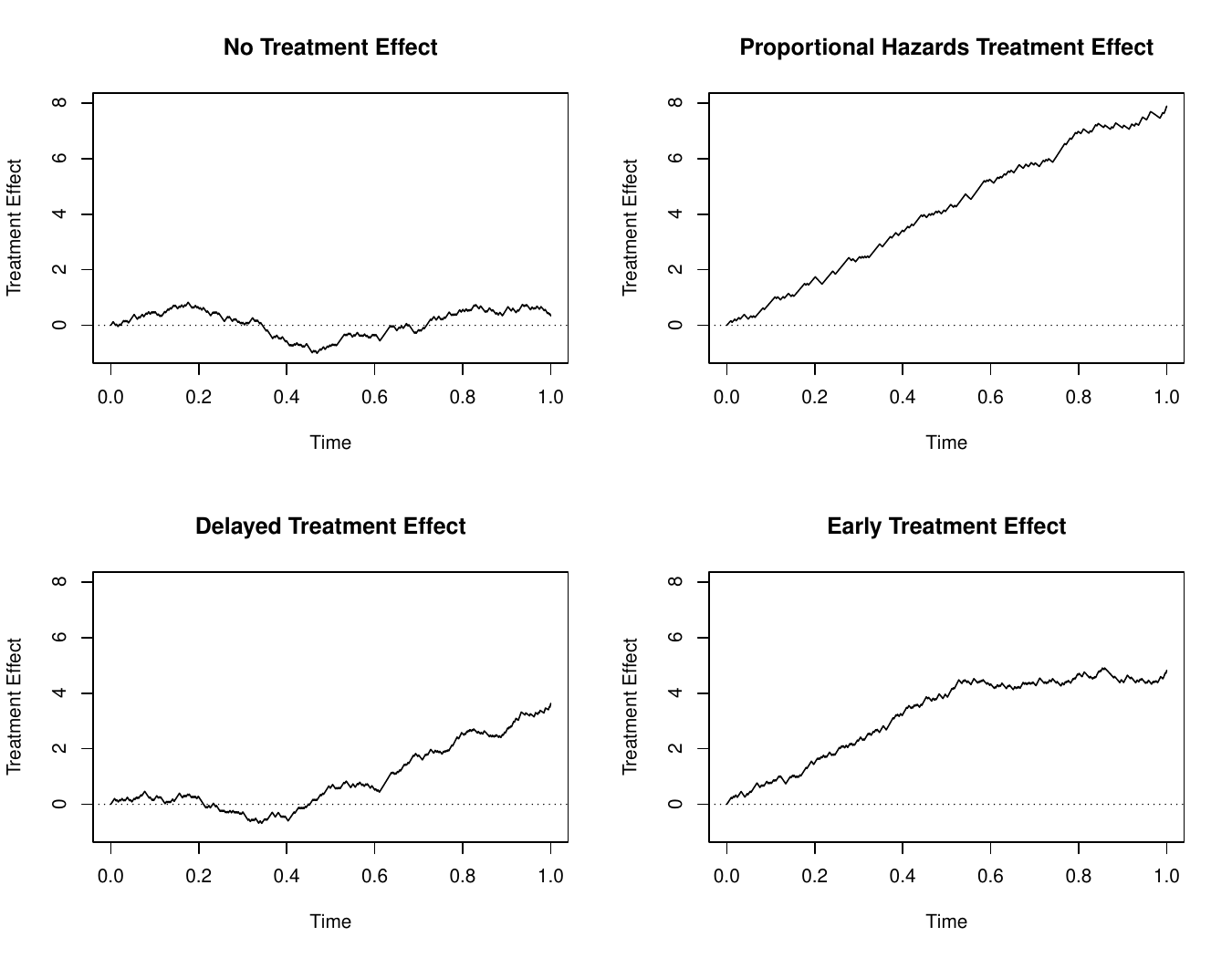}
 \caption{\label{fig.processes}The treatment effect process $U^*_n$ under different scenarios. The top left corresponds to no treatment effect, and the top right corresponds to a constant proportional hazards treatment effect. The bottom left and right correspond to a delayed treatment effect that occurs around the 40th percentile on the transformed time scale and an early treatment effect up until the 50th percentile.}
\end{figure}

We extend this to the whole $[0,1]$ interval by linear interpolation of the $k_n+1$ random variables $\left\{ U_n^*(\beta(t_j),t_j),\ j=0,1,\dots,k_n\right\} ,$ one of which is degenerate, taking on the value zero with probability one when $t_0 =0.$ We then have:   
the standardized score process, $U_n^*(\beta(t),t),\ t\in[0,1]$, evaluated at $\beta(t)$ where for $j=0,\dots,k_n$ and $t \in [t_j,t_{j+1}]$, is defined on (0,1) by:      
\EQ{
U^*_n\left( \beta(t),
t\right) = U^*_n\left( \beta(t_{j}), t_j\right) +
\left(tk_n-j\right) \left\lbrace U^*_n\left(\beta(t_{j+1}),
t_{j+1}\right)- U^*_n\left(\beta(t_{j}),t_j\right)\right\rbrace. 
}
Before establishing some needed theoretical results, we can gain some intuition from
Figure \ref{fig.processes} that describes a few situations. The first is the absence of effect, corresponding to Brownian motion. The other situations describe proportional and non-proportional hazards scenarios. 
The following theorem makes these observations more precise.
\begin{thm}\label{thNPH}
Let $\beta_0$ be a constant function on the class of possible regression functions $\mathbb{B}$. Under the non-proportional hazards model with parameter $\beta(t)$, consider
\begin{equation}
\label{CVderive}
A_n(t)= \dfrac{1}{k_n}{\displaystyle \int_0^{t}} \left\{\beta(s)-\beta_0\right\}\dfrac{\mathcal{V}_{\tilde \beta(s)}(Z\mid s)}{\mathcal{V}_{\beta_0}(Z\mid s)^{1/2}}d\bar N^*(s),\, \quad  0\leq t \leq 1,
\end{equation}
where $\tilde \beta $ is in the ball with center $\beta$ and radius $\sup_{t\in[0,1]}\left| \beta(t)-\beta_0 \right|$ and $N^*(s)$ is the counting process on the transformed time scale.
Then, there exist two constants $C_1(\beta,\beta_0)$ and $C_2 \in \R^{*+}$ such that $ C_1(\beta,\beta)=1$ and
\begin{equation}\label{derivebrownien}
 U^*_n(\beta_0,\cdot)-\sqrt{k_n}  A_n\quad \overset{\cal L}{\underset{n\rightarrow \infty}{\longrightarrow}}\quad  C_1(\beta,\beta_0) \mathcal{W}(t), 
\end{equation}  
where $\mathcal{W}(t)$ is a standard Brownian motion. 
 Furthermore,
\begin{equation} \label{CVderiveuniv}
 \sup_{0\leq t \leq 1} \left\vert A_n(t) - C_2\int_0^t\left\{\beta(s)-\beta_0\right\}ds\right\vert \quad  \overset{P}{\underset{n\rightarrow \infty}{\longrightarrow}}\quad   0.
\end{equation}
\end{thm}
\begin{cor} \label{TH_Null}
Under the non-proportional hazards model with parameter  $\beta(t)$, 
\begin{equation*}
U^*_n\left(\beta(\cdot),\cdot\right)\quad  \overset{\cal L}{\underset{n\rightarrow+\infty}{\longrightarrow}}\quad  \mathcal{W} (t).
\end{equation*}
\end{cor}
A proof of the theorem is given in \cite{Chauvel2014}. The theorem implies that the regression effect process will trace out a path that closely parallels that of $\int_0^t \beta (u) du .$ Simulations back this up, and it is straightforward to obtain curves such as those illustrated in Figure \ref{fig.processes}. On the original time scale, however, things will not look so clear cut so that, even in the case of proportional hazards, the process can appear to be non-linear. That said, on the finite set of time points, the transformation $\phi (t)$ is an isomorphism so when done with any calculation, we can always go back and forth between the original and the transformed time scales by employing $\phi (t)$ and $\phi^{-1} (\cdot ).$   

\subsection{The nph2ph-transform}

In practical situations, as opposed to theoretical study or simulations, the nph2ph transform will not be known. However, the treatment effect process, which is observed, provides direct information on the form of the transformation. As a consequence of some key results concerning the treatment effect process, we can construct a simple approximation to the unknown transform and assess to what degree this empirical transform has worked. We can be assisted in this by both visual and numerical quantification and clear guidelines on when to decide that our empirically derived transform is, in some sense, good enough. The following simple lemma is at the core of our development and states that every non-proportional hazard model, $\lambda(t|Z(t))=\lambda_0(t)\exp\{\beta(t) Z(t)\}$, is equivalent to a proportional hazards model.  
\begin{lmm}\label{thm.equivalence}
For given $\beta (t)$ and covariate $Z(t)$ there exists a constant $\beta_0$ and time dependent covariate $Z^* (t)$ so that  
$\lambda(t|Z(t))=\lambda_0(t)\exp\{\beta(t) Z(t)\} =\lambda_0(t)\exp\{\beta_0 Z^*(t)\}\, . $
\end{lmm}
There is no loss of generality if we take $\beta_0 = 1.$ The result is immediate upon introducing $\beta_0\neq 0$ and defining a time-dependent covariate $Z^* (t) \equiv Z(t) \beta (t)\beta_0^{-1}.$ The important thing to note is that we have the same $\lambda_0 (t)$ on either side of the equation and that, whatever the value of $\lambda (t | Z(t) )$, for all values of $t$, these values are exactly reproduced by either expression, i.e., we have equivalence. Simple though the lemma is, it has strong and important ramifications, and we continue to develop them here.

\section{Making improvements to the model}
The most common path to model improvement is by including regressors or predictor variables that improve the precision of our knowledge of the outcome variable. That path is well understood and not our focus here. Here, we are only concerned with ways to improve the model once we have already settled on the set of regressors or predictor variables. Since generalizations are quite straightforward, we only deal with the simplest case: a single binary regressor or treatment variable coded (0,1). Model improvements are then concentrated on the time effects. These improvements come under two closely related headings: goodness of fit and predictive strength.

\subsection{Improving model goodness-of-fit}\label{sec:improvefit}
There is a large body of literature focused on the problem of goodness-of-fit for proportional hazards models beginning with the test of \cite{Cox1972}. Many of these might be potential candidates for our purpose, although we will only focus on the approach of \cite{jmoq03}. That approach is based on the regression effect process, the treatment effect process in the case of a single binary variable, and so is fully in tune with the methods we use in this work. The regression effect itself, characterized by $\beta (t),$ need not be constant, which is also something we will need, which would rule out many other methods. Finally, our purpose is usually not to carry out any formal test with accurate control of error rates but to gain a visual impression of how well our model can be taken to accurately summarize time dependency. 

The visual summary we are looking for is based on the following. Under the model, the process $U^*_n\left( \beta(t), t \right)$
can be treated as though it were a realization of
a Brownian motion. Slutsky's theorem allows us to conclude that the same will hold for $U^*_n\left( \hat\beta(t), t \right)$ as long as  $\hat\beta (t)$ provides a consistent estimate of $\beta (t).$ Realistically, we will not be able to achieve that for general $\beta (t)$ but, for special cases, for instance, $\beta (t)$ constant or piecewise constant over a small division of the interval $(0,1),$ this can be manageable. 
From this we are able to derive the bridged process $U^0_n\left( \beta(t), t \right)$;  
\begin{defn}
The bridged process is defined by the transformation $$  U^0_n(\beta (t) , t)
 = U^*_n(\beta (t),t) -  t\times U^*_n(\beta (t),1) $$
\end{defn}
\begin{lmm}
The process $U^0_n(\beta (t) ,t)$ converges in distribution to the Brownian bridge,
in particular $E\, U^0_n\left( \beta(t), t \right) =0 $ and 
$\mbox{\rm Cov}\,\{U^0_n(\beta (s),s)$, $U^0_n(\beta (t),t)\}$ $=s(1-t).$
\end{lmm}
Once again, these results continue to hold for $U^0_n(\hat\beta (t) ,t)$ whenever $\hat\beta (t)$ is consistent for $\beta (t).$  
The Brownian bridge is also called tied-down Brownian motion for
the obvious reason that at $t=0$ and $t=1$ the process takes the
value 0. So when our model is ``correct'', i.e., it can be viewed as an
accurate representation of how the observations are generated, then this bridged process will
start and end at 0 and will show no obvious drift, appearing to walk randomly around the
origin axis. How far it may plausibly wander can be quantified by using: 
\EQ{\mbox{\rm Pr}\,\left\{ \sup_u | U^0_n(\hat\beta (t), u) |\ge a \right\} \approx 2\exp(-2a^2) , }
which follows as a large sample result from Kolmogorov's theorem; 
$$ \label{sigbrownbt}   \mbox{\rm Pr}\,\left\{ \sup_u |
U^0_n(\hat\beta (t), u) |\le a  \right\}   \to
1-2\sum_{k=1}^\infty (-1)^{k+1}\exp(-2k^2 a^2)\, , \ \ a \ge 0 \,
. $$  
This is an alternating sign series, and therefore, if we stop
the series at $k=2$ then the error is bounded by $2\exp(-8 a^2)$
which for most values of $a$ that we will be interested in will be
small enough to ignore.

Given the value of the process is 0 at $t=0$ and $t=1$, the variance of $U^0_n(\hat\beta (t), t)$ clearly depends on $t$. As an alternative, a standardized distribution, defined as $U^0_n(\hat\beta (t), t)/\sqrt{t(1-t)}$, could be utilized \citep{jmoq03}. This standardized distribution is not well defined at $t=0$ and $t=1$, so the interval (0,1) needs to be reduced to $(\epsilon_1, \epsilon_2)$, where $\epsilon_1 >0$ and $\epsilon_2<1$. With these components, we can use the following approximation by \cite{Miller1982},
\EQ{ \label{millersiegmund}   \mbox{\rm Pr}\,\left\{ \sup_u \left|
\frac{U^0_n(\hat\beta (t), u)}{\sqrt{u(1-u)}}   \right|\le a  \right\}   \approx \frac{4 \phi(a)}{a} +\phi(a) \left( a -\frac{1}{a}\right) \log \left\{ \frac{\epsilon_2(1-\epsilon_1)}{\epsilon_1(1-\epsilon_2)} \right\},}
where $a \ge 0$ and $\phi(x)$ is the standard normal distribution. With this result, we can use confidence intervals for the bridge process that go to 0 towards $t=0$ and $t=1$. When the bridged process strays into zones with small
probability under the large sample Brownian bridge assumption, this indicates that
the current model is not adequate. This will also be observable from the treatment effect process itself
and will simultaneously suggest a way for improvement. This can be seen in the examples below.  

\subsection{Improving predictive strength of the model} 
The predictive ability of a model is typically quantified by some $R^2$ measure of explained variation.
The reason for this follows as a consequence of Chebyshev's inequality showing that the smaller the variance, the more clustered, on average, the observations will be around the predicted conditional mean. The conditional mean is
all the more accurately located as the goodness-of-fit improves. And so the predictive strength of the model and
the goodness-of-fit, while quantifying different things, are closely related. We explore this further below, in particular,
using an important result of \cite{Chauvel2017} showing that the best fitting model will be the closest
to the model that generates the observations and that will maximize the population equivalent of our $R^2$ measure. 

We will measure predictive strength in two ways. We use $R^2$ that indicates the reduction in explained
variance due to the treatment indicator variable and, as a result of the Chebyshev inequality, the degree by which our
predictability has improved. Secondly, we use $\kappa,$ the concordance coefficient, that estimates how high the probability is that a random subject from the treatment group will live longer than one from the reference group. 

Note that, for any random variables $Z$ and $T$ having second moments, we have that  
$$\text{Var}(Z)=E(\text{Var}(Z\vert T=t))+\text{Var}(E(Z \vert T=t)).$$
From this the definition of explained variation, $\Omega^2,$ is immediate and we write,  
$\Omega^2 (\beta (t) ) = 1 -  \text{Var}(E(Z \vert T=t))  / \text{Var}(Z) =    E\text{Var}((Z \vert T=t))  / \text{Var}(Z)  .  $
In this expression we can replace the expectation operator $E$ by ${\cal E}$ and the variance operator $\text{Var}$ by ${\cal V}$ 
and, since these depend upon $\pi_i(\beta(t),t)$ which in turn depends upon $\beta (t)$ we make this clear, by writing $\Omega^2 (\beta (t)) ,$ emphasizing the dependence on the model as characterized by $\beta (t) .$ We also have:
\begin{lmm}
A consistent estimator of $\Omega^2 (\beta (t) )$ is $R^2 (\hat\beta (t) )$ in which $\hat\beta (t)$ is any consistent estimator of $\beta (t),$ in particular $\hat\beta,$ the usual partial likelihood estimator under proportional hazards. 
\end{lmm}
These results were shown in \cite{OQuigley1994}, where several properties were also outlined. These include large sample independence of $R^2$ to an independent censoring mechanism, invariance to linear transforms, and monotonically increasing transforms on time, i.e., rank invariance. For our purposes, we need more, and the following results support our development. Suppose that 
$\mathcal{B}$ defines a class of possible regression functions $\beta (t).$ This class can be as large as we wish, restricted in any way we wish, i.e., monotonic increasing or decreasing, and not necessarily continuous. It does, though, have to be finite. Then: 
\begin{thm}
\label{theoremR2}
Under the non--proportional hazards model (\ref{nonph}) of parameter $\beta (t)$, 
we have $$\vert R^2(\beta (t))-R^2(\hat \beta (t)) \vert\overset{a.s.}{\underset{n\rightarrow + \infty}{\longrightarrow}} 0,$$
and with probability one,
$\underset{{b (t) \in \mathcal{B}}}{\arg\max}\ \lim_{n\rightarrow + \infty} R^2(b(t))=\beta (t).$ 
\end{thm}
The proof is given in Chauvel and O'Quigley (2017). If $\beta (t)$ is the true regression coefficient, the maximum of $R^2$ is then well approximated by $R^2(\hat \beta (t))$ for a large enough sample size. 
Using the theorem results, the treatment effect process can indicate the form of the regression effect in relation to time. Tools such as smoothing or the projection on a basis of functions or kernel estimation allow for other approaches
\citep{Cai2003,Hastie1990,Scheike2004}. We have not compared the different tools. 
As an illustration, Figure \ref{fig.processes} shows a constant effect until time $\tau$ followed by a null effect. This is easily detectable, especially with moderate and larger sample sizes.

We can assume that $\beta (t)$ can be written $\beta_0 \times b (t)$ where $b(t)$ is a member of $\mathcal{B} .$ Given $b (t)$ we replace $\beta (t)$ by $\hat\beta (t) = \hat\beta_0 \times b(t)  $ and $\hat\beta_0$ is obtained in any of the usual ways, eg via maximization of the partial likelihood \citep{Cox1972}. The function $ b (t)$ can be determined graphically from the treatment effect process, noting that this function is only specified up to an arbitrary constant. In a single changepoint model, all that is needed to obtain $b (t)$ is the ratio of the two slopes. More formally,     
let $\mathcal{B}=\{b_1 (t),\dots,b_m (t)\}$ be a set of
$m$ functions from $[0,1]$ to $\R$. Note that while $m$ is not unbounded, it can be as large as we wish.   
The selected regression function $b^* (t)$ is such that
$$ b^* (t)=\arg\max_{b(t) \in \mathcal{B}}R^2\left(b(t)\right)  .$$
The following theorem gives an equivalence between this maximization problem when $n \rightarrow \infty$ and a problem of minimization of $L^2$ norm.
\begin{thm} 
\label{theoremR2L2}
Under the non--proportional hazards model with regression parameter $\beta(t)$ not necessarily in $\cal B$, asymptotically, $ \beta^*(t)=\underset{\alpha(t) \in \mathcal{B}}{\arg\max}\ \underset{n\rightarrow \infty}{\lim}R^2\left(\alpha(t)\right)$ is the solution of
$$ \beta^*(t)=\arg\min_{\alpha(t) \in \mathcal{B}} \left\| \beta(t)-\alpha(t)\right\|_{2}=\arg\min_{\alpha(t) \in \mathcal{B}} \left(\int_0^1\left( \beta(t)-\alpha(t)\right)^2dt\right)^{1/2}.$$
\end{thm}
Other approaches may be possible but we have not explored them in view of a lack of results regarding their properties.
Several estimators of the explained variation have been proposed in the literature \citep{Choodari2012}. Here we have avoided presenting an exhaustive review of these estimators. Instead, we have focused on the $R^2$ coefficient introduced by \citet{OQuigley1994} since it is built with the same ingredients as the treatment effect process. Our proposal in this paper leans on several well-established results.  
 
 \section{Working approximations for the nph2ph-transform}\label{sec.approxp}
Our goal is not to find the best possible estimate of the nph2ph transform, a goal that would be very challenging both conceptually and operationally. Instead, we aim to find a simple approximation to the nph2ph transform that appears to do well enough. Our definition of ``well enough'' equates to a transform that results in a bridged process that does not stray outside the boundaries established according to Equation (\ref{millersiegmund}), i.e., in Figure \ref{fig.longNEJM} (D) for an upcoming example, one that lies fully inside the unshaded zone. According to \cite{Chauvel2017}, once such a situation prevails,, there is very little room for increasing the model's predictability unless, of course, we add in new or further covariates. 
Briefly, if some simple approximation to nph2ph brings a bridged process, some of which has strayed into the shaded areas, into one that is fully contained within the unshaded area, then we can claim an improvement of the proportional hazards approximation and anticipate a resulting improvement in predictive power. This has been the case with the examples shown here, and with many others, we have looked at. 

While the treatment effect process conveys the needed information on the shape of the nph2ph transform, it also contains an amount of irregular noise that we may wish to eliminate. Drawing a well-fitting curve over the process would provide a working solution but would not be readily reproducible. Most likely, there are many possible solutions provided by any established curve fitting technique. In the next two sections, we describe two of these, both of which we exploit in our examples. The first solution provides a smooth, continuously differentiable curve. The second is a rough and ready, albeit powerful, approximation based on a change-point model.   

\subsection{Legendre polynomials with bounded second derivatives}\label{sec.orthpoly}
Recall that we aim to find something that is ``good enough'' for the data at hand, meaning that we want some working summary of what has been observed. Orthogonal polynomials can be a useful way to approximate functions. Their accuracy will depend on the number of included terms, and, in our situation, they have some appeal in that such functions are both continuous and differentiable. They are no more than working approximations, and although many orthogonal bases arise as solutions to differential equations of mathematical physics, any attempt to make physical connections might not seem realistic. That said, orthogonal polynomials will allow us to build our approximations incrementally so that the higher the order, the better the job they do. There is, though, a downside to that: the problem of overfitting, whereby a great deal of flexibility can result in random events and background noise having too strong an influence on the fit. We can readily control this by not allowing the fitted curve to turn too quickly, i.e., by putting bounds on the second derivative.   
One candidate class of orthogonal polynomials that comes to mind is the class, $P_0 (t),$ $P_1 (t), ... ,$ known as the Legendre polynomials. We have that $P_0 (t) =1$ and $P_1 (t) = t$ so that these provide the basis for the absence of regression effect and a proportional hazards effect. Higher orders obtained from
$$  (n+1 )\times P_{n+1} (t) = (2n + 1)\times t P_n (t)  -   n\times P_{n-1} (t) \, , 
$$
bring steadily improving fits. In our limited experience, it will generally suffice to employ no more than lower-order Legendre polynomials. One helpful fact is that the derivatives, $\Delta (P_n)$, of Legendre polynomials can be expressed in terms of polynomials of lower order. This enables us to obtain a visual impression of the form of $\beta (t).$ In Figure \ref{fig.longNEJM}, we see a Legendre approximation and a piecewise PH approximation, described just below. Both offer major improvements over the simple PH model.

\subsection{Piecewise proportional hazards models}\label{sec.pphm} 

A piecewise PH model with a single changepoint would be the first step in approximating the nph2ph transform. Indeed, in the examples we show and many others we have looked at, this very simple first step away from PH can be very effective. Given real observations, there are many ways in which such a model could be fit. Among these would be the changepoint survival models described by \cite{OQuigleyPessione1991}, \cite{Contal1999}, and \cite{ OQuigleyNatarajan2004}. We could also focus our attention on the treatment effect process itself, which, under a simple assumption of a two-stage PH model, would provide two regression slopes to be estimated. The technique of \cite{Chow1960} is a natural contender for this estimation although it would not guarantee continuity of the slopes at the changepoint and, given that our goal is to model Brownian motion with drift, this would seem like a weakness. Our preferred approach is to use Euler's theorem for the drift parameter of Brownian motion with linear drift. This provides us with two regression slopes that are continuous everywhere although, of course, differentiable nowhere in a large sample sense, as indeed is Brownian motion itself.  
We will appeal to known theoretical results of Brownian motion and the Brownian bridge. 

Let us take the changepoint ${\tau}$ to be the point at which the slope changes and our estimate of this is based on optimizing fit so that; $\tau = \arg \max_{i=1}^{k_n}  L (s ) $ where;
$$ L (s )  =   \sum_i \left\{ \log   \pi_i(\beta_1 (X_i), X_i) I(X_i <s)  +  \log\pi_i(\beta_2 (X_i), X_i) I(X_i \geq s)     \right\}   .    $$
Now that we have our estimate for $\tau ,$ we can write down the piecewise linear function $ U^P_n (\beta (t), t )$ that best fits the observations. Note that by best fit, we mean the function consequent upon maximizing the likelihood $L (s)$ across all possible changepoints. Given any changepoint, we only need to estimate the drift coefficient of Brownian motion for before and after $\tau .$ We do this by appealing to Euler's theorem whereby a simple linear connection of the first point on the process to the last (on either side of $\tau$) would correspond to the maximum likelihood estimate of the drift parameter for Brownian motion with linear drift. We then have;
$$   U^P_n (\beta (t), t )  =    \frac{  t\times U^*_n (\beta (t), \tau  ) \La (t<\tau)   }{ \tau }
                             +    \frac{  [ (t -\tau )  \times  U^*_n (\beta (t), 1  ) + U^*_n (\beta (t), \tau  ) (1 -t  )   ]  \La (t\geq \tau )   }{1-\tau}  
$$
We can see that this function is piecewise linear and, for our purposes, we need take note of two slopes, the one before and the one after the changepoint $\tau .$ This provides us with the form for $\beta (t)$ which for computational purposes we will re-express in terms of $Z(t).$ Consider then, 
\begin{lmm}
The piecewise proportional hazards model with time-dependent coefficient $\beta (t)$ and changepoint $\tau$ on the transformed scale is equivalent to a proportional hazards model with time-dependent treatment indicator, $Z^P(t)$ in which;
\EQ{\label{piecez}  Z^P(t) =  \La (t\leq\tau) \times Z(t) +   \frac{\tau \La (t> \tau) }{ 1 - \tau } \left\{ \frac{U^*_n (\beta (t), 1  ) -  U^*_n (\beta (t), \tau  ) }{  U^*_n (\beta (t), \tau  )  }    \right\} \times Z(t)
}
\end{lmm}

\begin{pro} 
\label{pro.r2g}
Suppose the observations are generated by the NPH model with coefficient function $\beta(t).$ The set ${\cal B}_1$ is the collection of all proportional hazards models and the set ${\cal B}_2 ,$ the collection of all piecewise PH models with a changepoint at $\tau .$ Then $E R^2 (\beta_2 (t) ) \geq E R^2 ( \beta_1   )$ where:
$$ \beta_j (t) =\arg\min_{\alpha(t) \in \mathcal{B}_j } \left(\int_0^1\left( \beta(t)-\alpha(t)\right)^2dt\right)^{1/2}\, , \ \ \ j=1,2.$$
\end{pro}
The proposition is simply a corollary to Theorem \ref{theoremR2L2} and is immediate. It tells us that our predictive strength, as measured by $R^2 ,$ will not decrease in expectation when we replace the simple proportional hazards model by the single changepoint piecewise proportional hazards model. Again, this assumes that we do not change the covariates in the model and how they are identified. We may, of course, purely for computation, re-express a time-dependent effect by a time-dependent covariate as in Equation \ref{piecez}. No optimality properties are claimed for the transform, and its success is to be judged by its impact on fit and the improvement in predictive strength. 

Equation \ref{piecez} looks a little involved at first glance but is no more than a time-dependent rescaling of the covariate based on the ratios of the two slopes of the process. If these slopes are the same, then the covariate is unchanged. Note also that any PH model remains a PH model under a linear transformation of the covariate. The product of the regression coefficient, $\beta$, and the covariate, i.e., the log relative risk, does not change, and so, in Equation \ref{piecez}, the covariate value before $\tau$ can be taken to be the original coding, $Z(t).$ We can use this observation to find the piecewise approximation to the nph2ph transform.

\section{Coefficient of concordance}
For our purposes, we make use of $R^2$ in the light of its many desirable properties, in particular, the properties described in the above subsections, allowing us to build better-fitting models from poorer-fitting ones. Many other coefficients, similar in spirit to $R^2$ have been suggested in the literature, see for example \cite{Harrell1996, Gonen2005, Uno2011, Choodari2012}, and these might also serve the purpose of our choice here. One logical choice since it directly translates predictability is the coefficient of concordance, $\kappa (A,B) .$ Recall that, for $R^2$ and two treatment groups, $A$ and $B$ with survival times $T_A$ and $T_B,$ $R^2$ will tend to the value zero when the group rankings are random, unrelated to the group status, and will tend to the value one when the sets of rankings become almost perfectly separated. Moving away from independence, i.e., $R^2=0,$ we know that, in expectation, $R^2$ increases as the degree of concordance increases  \citep{Xu1999} and we measure concordance via;
$$  \kappa (A,B)= 1 - \kappa (B,A) = \mathrm{Pr}\, (T_A > T_B)
$$
\begin{lmm}\label{lmm.concbeta} 
Suppose that $T_A$ relates to $T_B$ via a proportional hazards model with parameter $\beta ,$ then:
  $$      \kappa (A,B) =  \frac{1}{  1+ \exp ( - \beta )     }   $$
\end{lmm}
When $T_A$ and $T_B$ are related via the simple exponential model, we can obtain the result by the convolution of $T_A - T_B$ and simple integration. The result then immediately extends to more general proportional hazards models. A useful measure of predictive strength is the relative risk $\psi (A, B)$ where $\psi (A,B) = \kappa (A,B)  /  (1- \kappa (A,B)  ) . $ When $\kappa (A,B) = 0.5 ,$ there is no regression effect and $\psi (A,B) = 1.$ Suppose that $\psi(A,B) = 1.5 .$ We could then say that under treatment, $A, $ a patient has a 50\% greater odds than one under treatment $B$ of being the longer survivor of the two. The most complex model we consider here is a model with a single changepoint, $\tau.$

\begin{lmm} For a piecewise PH model and parameters $\beta_1 (T<\tau)$ and $\beta_2 (T>\tau)$ then; 

\begin{eqnarray*}
  \kappa (A,B) =  \frac{\{  1 - \exp (- \tau)  \}   \{1-\exp ( -\tau \beta_1 )  \}   }{   1+ \exp (-\beta_1) }   
   +    \frac{ \exp \{ - \tau(1+ \beta_2 ) \}  }{1+\exp(-\beta_2)}   +    \exp (-\tau )  \{ 1- \exp (-\tau\beta_1)    \}
  \end{eqnarray*}  
  
\end{lmm}
\noindent Proof: A simple partition of the outcome space leads to; 
\begin{eqnarray*}
\kappa (A, B)  =  
&=&  \Pr(T_A > T_B | T_A < \tau , T_B < \tau  ) \times \Pr(T_A < \tau , T_B < \tau  ) + \\
&&  \Pr(T_A > T_B  | T_A > \tau , T_B < \tau   ) \times\Pr( T_A > \tau , T_B < \tau    ) + \\
&&  \Pr(T_A > T_B  | T_A < \tau , T_B > \tau   )\times\Pr( T_A < \tau , T_B > \tau  ) + \\
&&  \Pr(T_A > T_B  | T_A > \tau , T_B > \tau  )\times \Pr(  T_A > \tau , T_B > \tau  )   .  
\end{eqnarray*}
Using Lemma \ref{lmm.concbeta}, standard results of the exponential distribution in conjunction with the rank invariance of proportional hazards models, the independence of survival times, the simplification that $\Pr(T_A > T_B  | T_A > \tau, T_B < \tau  )=1$ and $\Pr(T_A > T_B  | T_A < \tau, T_B > \tau  )=0$, and gathering together all of the terms, we have the result. While we can see that the details can quickly become fastidious, it is clear that the extension to several change-points raises no conceptual difficulties.

\section{Using the approximated nph2ph transform to estimate the conditional survival functions}
It is common to estimate the survival functions via non-parametric methods such as Kaplan-Meier or Nelson-Aalen since parametric models, although relatively easy to fit, will rarely fit well to real data. A proportional hazards model in the case of say two groups allows us to obtain estimates that are similar in spirit but that share information. The groups are not independent. The estimated regression coefficient of the model will fully capture the dependency structure of the two groups. In the case of proportional hazards the two survival functions will look like power transforms of one another. This relationship breaks down for non-proportional hazards situations, and the PH fitted models may do a poor job in approximating the survival curves. This is well illustrated by Figure \ref{fig.LongIntro} of the Introduction. The nph2ph transform is a very effective tool that facilitates the estimation of the survival function in NPH situations.   

The properties of the Nelson-Aalen estimates are asymptotically equivalent to the Kaplan-Meier estimates. In practical situations, the two estimates tend to all but coincide, and, in our view, the Nelson-Aalen estimate is a better choice for the development since the relationship to PH and NPH models is more transparent. To see this, for two groups, $G=0$ and $G=1,$ the survival functions  given by $S_G (t) = \exp \{- \Lambda (t:G) \}\, , G=0,1,$ in which, 

\begin{eqnarray} \label{conditionalsurv}
 \Lambda (t:G)  =\, \sum_{i=1}^n   \frac{  \La( {X_i \leq t} )  \delta_i  \exp\{\beta(X_i ) G \} }{ \sum_{j=1}^n Y_j(X_i) \exp\{\beta(X_i) Z_j(X_i)\} }, \text{ }
\end{eqnarray}

\begin{figure}[h]
\centering
 \includegraphics[width=3in, height=5in]{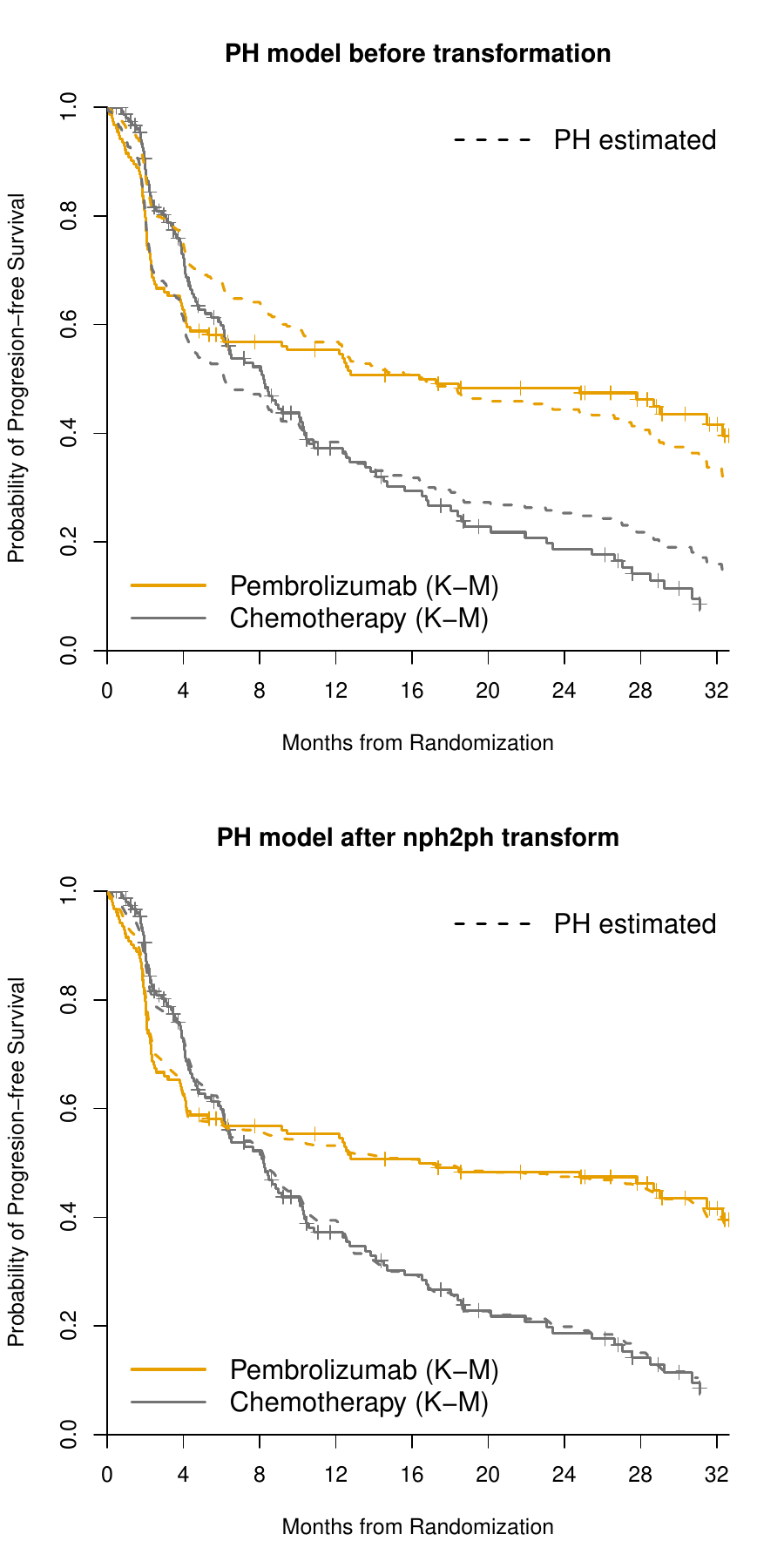}
 \caption{Kaplan-Meier curves from a survival study in colorectal cancer \citep{Andre2020}. The top figure shows the estimated conditional survival functions obtained under the best-fitting proportional hazards model. The bottom figure shows the estimated conditional survival functions under the best-fitting proportional hazards model after implementing an nph2ph transform.  \label{fig.Andre2020supp}   }
\end{figure}

When we have a single group, or when $\beta (t) = 0,$, then this expression coincides with the usual Nelson-Aalen estimate, which, in turn, all but coincides with the Kaplan-Meier estimate. 
In practice we replace $\beta (t)$ by $\hat\beta  (t).$ Not precisely knowing $\beta (t)$ implies that we do not know precisely nph2ph. However, the most elementary approximation to nph2ph can work very well. We saw this in the example of the Introduction. We have seen it in other examples, but the lack of space prevents us from showing it here. For illustration, we consider one further example in which it is immediately clear that PH provides a poor model, while the most elementary of nph2ph approximations, a single change point model, indicates a significantly improved fit. The improvement, as expressed via the conditional survival function in Equation (\ref{conditionalsurv}), is so strong that any further improvements will necessarily be very small (see Figure \ref{fig.Andre2020supp}). This example is further discussed in the following section.

\section{Recent examples from the literature}  
The nph2ph transform is demonstrated using multiple published phase III trials in oncology. The trials were selected to illustrate different types of treatment effects observed across various clinical settings. As the individual patient-level data were not directly available for each trial, the published survival curves were digitized to recreate individual-level data for this illustration \citep{Guyot2012,Satagopan2017}.

In each randomized trial, a proportional hazards regression model estimated the hazard ratio associated with each treatment effect. The goodness-of-fit was assessed for each model by examining the bridged treatment effect process outlined in Section \ref{sec:improvefit}. Confidence bounds were added to each bridge plot using Equation (\ref{millersiegmund}) to aid model assessment. Two sets of bounds were added to each figure (see, for example, plot (B) in Figure \ref{fig.longNEJM}). If the bridged process were to stay within the center white area, it would indicate the current model provides a reasonable fit. The border of the light gray area represents a 90\%  confidence bound for the bridge, and the dark gray area represents a 99.9\% confidence bound; if the bridged process were to enter these shaded areas, it would represent different degrees of model inadequacy. 

A Shiny app of the nph2ph transform is available to confirm these results or run on additional data sets (https://nph2ph.shinyapps.io/nph2ph/). 

\subsection{Dabrafenib and Trametinib Compared to Placebo in Patients with Melanoma Cancer}\label{sec.trialdab}
\cite{LongNEJM2017} conducted a phase III randomized trial to compare combination therapy with dabrafenib and trametinib to a placebo in melanoma patients with a BRAF mutation.  The trial enrolled 870 patients who were randomized 1:1 to the two treatment arms. The primary endpoint of the study was relapse-free survival. At the time of trial reporting, a total of 414 patients died or relapsed.

\begin{figure}[h]
\centering
 \includegraphics[width=6.0in, height=7.0in]{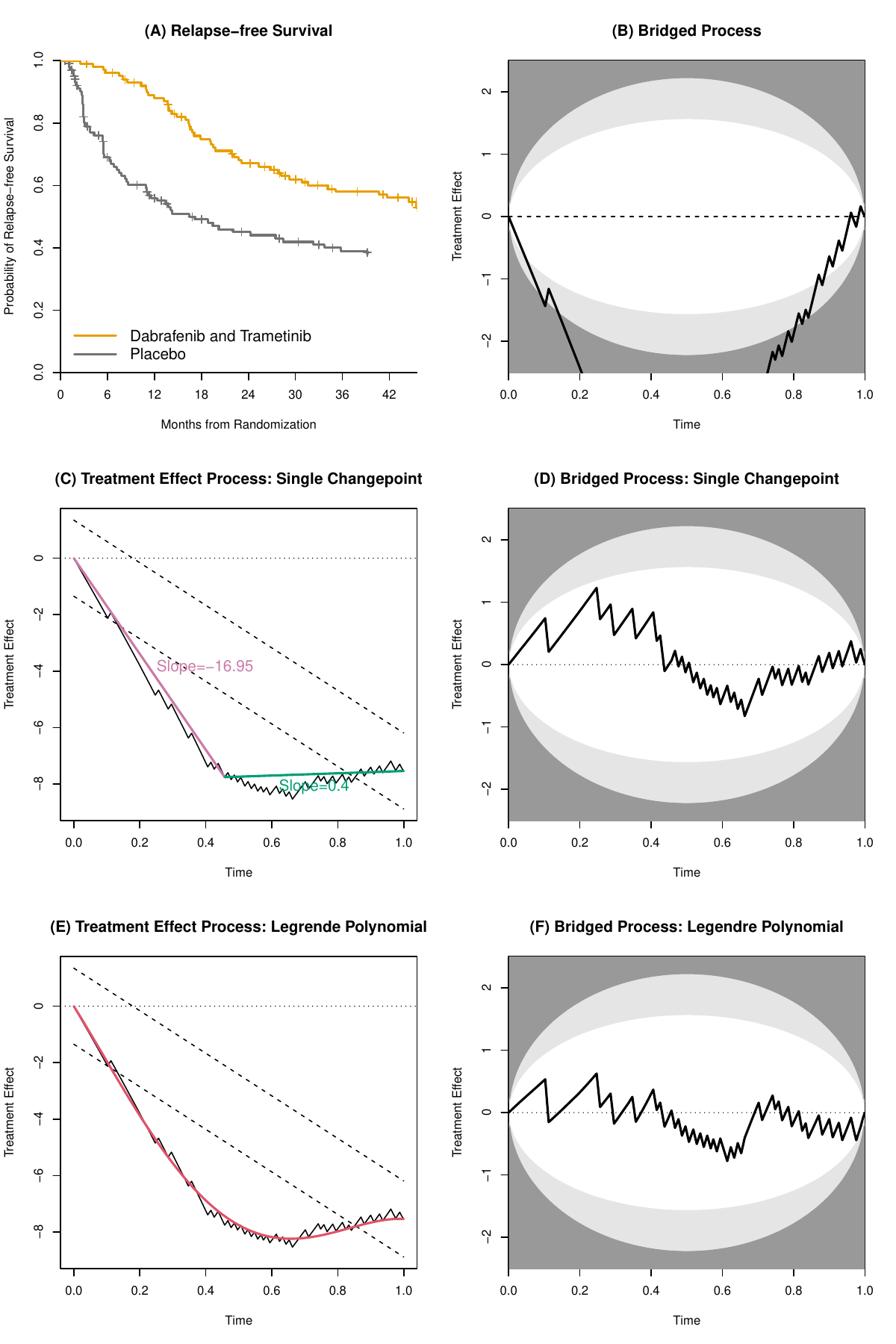}
 \caption{\label{fig.longNEJM} (A) Estimated relapse-free survival for patients treated with dabrafenib and trametinib compared to placebo in \cite{LongNEJM2017}. (B) The bridged treatment effect process indicating a major model deviation. (C) The treatment effect process with the identified changepoint $\tau=0.46$ and the estimated slopes before and after $\tau$. (D) The bridged process after the implementation of the changepoint. (E) and (F) are the estimated Legendre polynomials for the treatment effect process and the subsequent bridged process. }
\end{figure}

The estimated Kaplan-Meier relapse-free survival curves are shown in Figure \ref{fig.longNEJM} (A) for the digitized data set. Using proportional hazards regression, the estimated hazard ratio was 0.47 (95\% CI: 0.38-0.57). The $R^2$ was modest at 0.133. 

Investigation of the bridged regression process in Figure \ref{fig.longNEJM} (B) indicates a major deviation from proportional hazards. The maximum of the bridged process was 4.45, which, using the bounds outlined in Equation (\ref{millersiegmund}), indicates a very poor model fit.

Using the treatment effect process provided in Figure \ref{fig.longNEJM} (C), a changepoint was identified at transformed time $t= 0.46$, which corresponds to approximately 8.5 months on the original time scale. The ratio of the two estimated slopes shown in the figure before and after the changepoint is -0.024. This results in an estimated effect of $\hat{\beta}(t)= -2.08 \times (I\{t < 0.46\} - 0.024 \times  I\{t \geq 0.46\})$. Interpreting these results, there is a large estimated effect on the rate of relapse or death when treated with dabrafenib and trametinib compared to placebo in the first eight months following randomization with an estimated hazard ratio of 0.12. However, after eight months, the estimated effect largely disappears with a hazard ratio of 1.02. 

The $R^2$ for this new model increased considerably to 0.306.  The bridged process for this new model is shown in Figure \ref{fig.longNEJM} (D). The bridged process remained within the denoted bounds; therefore, there was little indication of model inadequacy after implementation of the single changepoint. With this new model well approximated by proportional hazards, the concordance coefficient was estimated to be 0.668.

Despite having no indication of model misspecification after implementing a single changepoint, a two-changepoint model could be explored, given the severity of proportional hazards deviation in the original data. This richer model resulted in a modest improvement to fit, with an improvement in $R^2$ from 0.306 to 0.327. Collectively, this indicates that a single changepoint model on its own was enough to overcome the large model misspecification that was observed when the original data were fit using proportional hazards regression.  

 An alternative approach to approximating the nph2ph transform could be based on the Legendre polynomials as outlined in Section \ref{sec.orthpoly}. One such polynomial is provided in Figure \ref{fig.longNEJM} (E). The bridged process in the next panel indicates a reasonable model fit. The resulting $R^2$ for this polynomial is 0.317, a modest improvement over the single changepoint model. The estimated effect under this approximation is $\hat{\beta}(t)= 1.50 \Delta(P_1(t))-3.06 \Delta(P_2(t))+2.62 \Delta(P_3(t))-0.81\Delta(P_4(t))$.
    
\subsection{Pembrolizumab Compared to Chemotherapy in Patients with Colorectal Cancer with High Microsatellite Instability} 

\cite{Andre2020} conducted a phase III randomized study of pembrolizumab compared to chemotherapy among advanced or metastatistc colorectal cancer patients who have tumors with high microsatellite instability or that are mismatch repair deficient. A total of 153 patients were randomized to receive pembrolizumab and 154 were randomized to receive chemotherapy. The primary endpoint was progression-free survival; the study had a median follow-up of 32.4 months.  

\begin{figure}[h]
\centering
 \includegraphics[width=6.0in, height=7.0in]{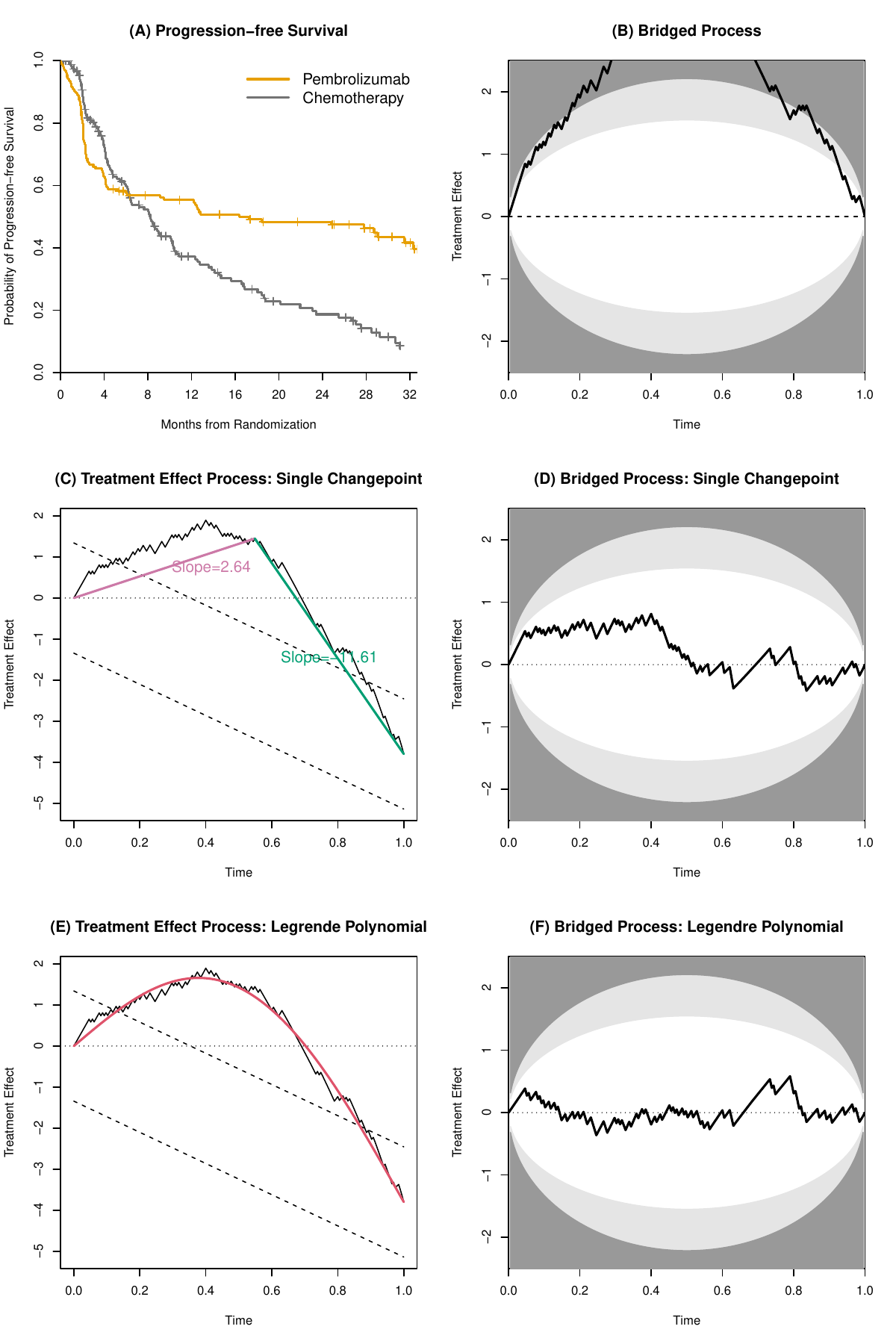}
 \caption{\label{fig.AndreNEJM} (A) Estimated progression-free survival for colorectal cancer patients treated with pembrolizumab compared to chemotherapy in \cite{Andre2020}. (B) A major proportional hazards model deviation was identified from the bridged process. (C) The treatment effect process with the identified changepoint $\tau=0.55$ and the estimated slopes before and after $\tau$. (D) The bridged process after the implementation of the changepoint.  (E) and (F) are the estimated Legendre polynomials for the treatment effect process and the subsequent bridged process.  }
\end{figure}

Figure \ref{fig.AndreNEJM}(A) shows the Kaplan-Meier survival curves for the two treatment groups. From these curves, there appears to be a deviation from the proportional hazards, a deviation we can also observe from Figure \ref{fig.AndreNEJM}(B) and (C). The maximum of the bridge was 3.51. In the digitized data set, the hazard ratio was estimated to be 0.59 (95\% CI: 0.44-0.79) and the $R^2$ was 0.053. 

\cite{Andre2020} recognized the non-proportionality of the data and supplemented the proportional hazard regression results with an analysis of the restricted mean survival time. We include this example to illustrate how the nph2ph transform can be used as a straightforward approach to interpret the treatment effect in this setting. 

A changepoint of 0.55 was identified using the treatment effect process in Figure \ref{fig.AndreNEJM}(C). On the original time scale, the changepoint corresponds to 4.2 months. Using the ratio of -4.40 for the two estimated slopes, the estimated treatment effect after implementing the changepoint was $\hat{\beta}(t)= 0.41 \times (I\{t < 0.55\} -4.40 \times  I\{t \geq 0.55\})$, which shows the benefit of pembrolizumab after the changepoint. The $R^2$ of the new model increased considerably to 0.265.  

The improvement in fit is also illustrated by examining best-fitting proportional hazards models before and after nph2ph implementation, as shown in Figure \ref{fig.Andre2020supp}. A clear improvement is observed.  

Similar to the previous section, Legendre polynomials were explored as an alternative nph2ph transform. This is shown in Figure \ref{fig.AndreNEJM}(E), and the bridged process is shown in Figure \ref{fig.AndreNEJM}(F). There appears to be no evidence of model inadequacy.  The estimated effect under this approximation is $\hat{\beta}(t)= -0.98\Delta(P_1(t))+1.00\Delta(P_2(t))-1.32\Delta(P_3(t))+0.38\Delta(P_4(t))$. The $R^2$ for this approach was 0.274, a value close to 0.265 based on the changepoint model.

\subsection{Cetuximab Compared to Best Supportive Care Patients with Colorectal Cancer}
A phase III randomized study compared cetuximab to best supportive care for patients with advanced colorectal cancer expressing EGFR \citep{Jonker2007}. A total of 287 patients received cetuximab, while 285 received best supportive care. The investigators found that cetuximab significantly improved both overall survival and progression-free survival. 

\begin{figure}[h]
\centering
 \includegraphics[width=6.0in, height=7.0in]{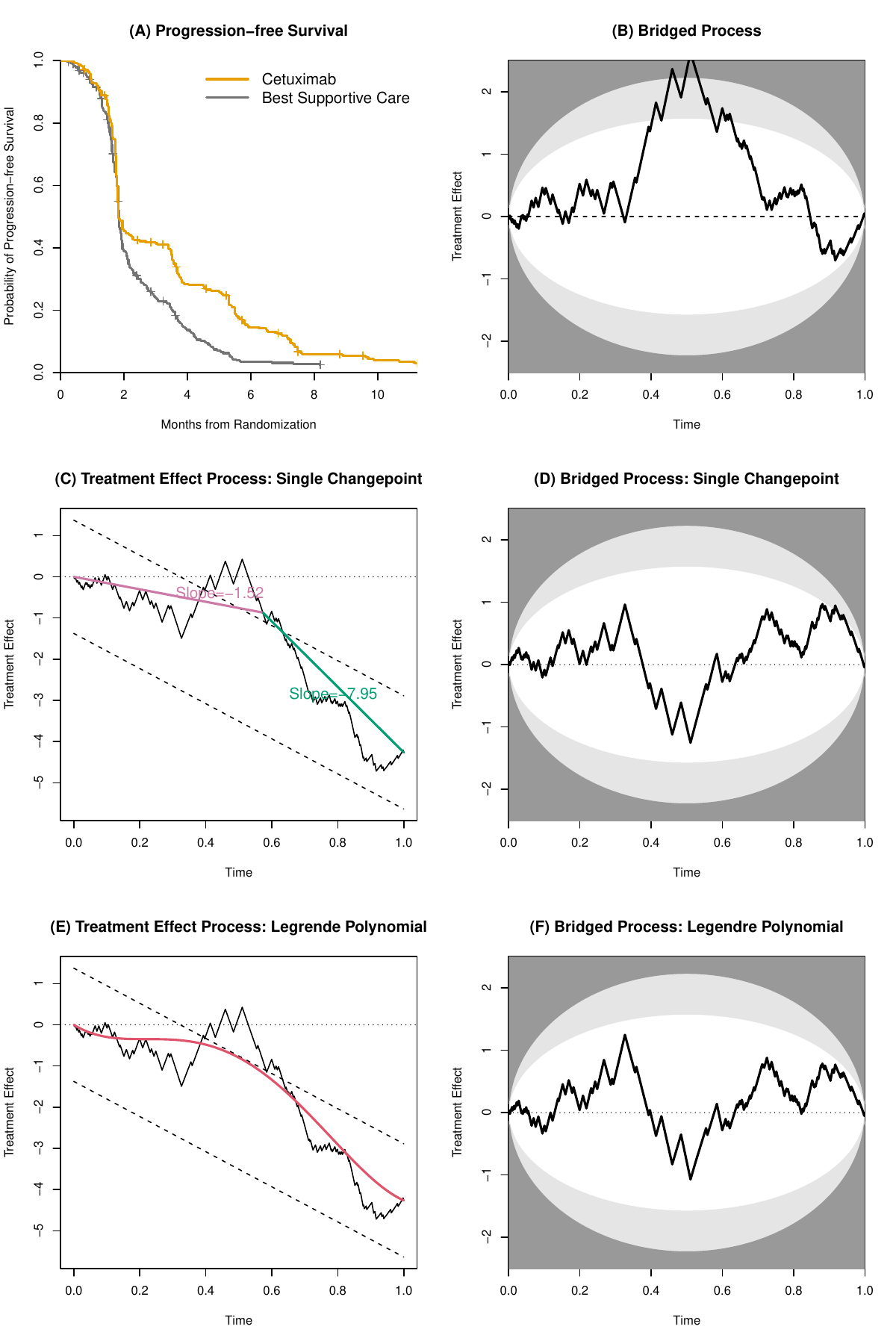}
 \caption{\label{fig.JonkerNEJM} (A) Estimated progression-free survival for patients treated with cetuximab compared to best supportive care in \cite{Jonker2007}. (B) The bridged treatment effect process indicates a deviation from proportion hazards with a maximum value of 2.63. (C) The treatment effect process with the identified changepoint $\tau=0.57$. (D) The bridged process after the implementation of the changepoint was within the denoted boundaries. (E) and (F) are the estimated Legendre polynomials for the treatment effect process and the subsequent bridged process. }
\end{figure}

The progression-free survival curves for the two treatment arms are shown in Figure \ref{fig.JonkerNEJM} (A) for the digitized data set. Visually, from the Kaplan-Meier curves, there is a clear delay in the treatment effect occurring around two months following randomization.  The estimated hazard ratio for the treatment effect was 0.69 (95\% CI: 0.58-0.82) using proportional hazards regression. The $R^2$ was 0.03.

The bridged process, shown in Figure \ref{fig.JonkerNEJM} (B) indicates deviation from proportional hazards with a maximum of 2.63. Using the regression effect process in (C),  the nph2ph transform identified 0.57 as the changepoint on the transformed time scale, corresponding to 1.92 months on the original time scale. The ratio of the two estimated slopes before and after the changepoint is 5.24. This results in an estimated treatment effect on the transformed time scale of $\hat{\beta}(t)= -0.089 \times (I\{t < 0.57\} + 5.24 \times  I\{t \geq 0.57\})$. The hazard ratios of 0.91 and 0.63 before and after the changepoint at 1.9 months from the nph2ph transform align with what is visually observed in the Kaplan-Meier curves.  

The $R^2$ for the new model is 0.051. The bridged process in Figure \ref{fig.JonkerNEJM} (D) indicates no further model deviation after implementing the single changepoint. The concordance is estimated to be 0.558 for the new model.

The alternative polynomial approach to nph2ph and the corresponding bridged process are shown in Figure \ref{fig.JonkerNEJM} (E)-(F).   The estimated effect for this approximation is $\hat{\beta}(t)= -3.54\Delta(P_1(t))+3.17\Delta(P_2(t))-2.01\Delta(P_3(t))+0.58\Delta(P_4(t))(t)$. The $R^2$ for this estimate is 0.077. Interestingly, increasing the order of the Legendre polynomial to $P_8(t)$ only results in an $R^2$ improvement to 0.096 despite aligning more closely with the treatment effect process. This supports the idea that once the bridged process is within the bounds in Figure \ref{fig.JonkerNEJM} (F), there will be only a marginal $R^2$ gain in further developing a richer model. 

\subsection{Ipilimumab Compared to Placebo in Patients with Melanoma}
In the final example, Eggermont and colleagues conducted a phase III randomized trial of ipilimumab compared to placebo in patients with stage III melanoma \citep{Eggermont2016}. A total of 951 were randomized 1:1 to the two arms. Over a five-year median follow-up of 5.3 years, the authors found that ipilimumab significantly improved overall and recurrence-free survival.  

\begin{figure}[h]
\centering
 \includegraphics[width=5.0in, height=5.0in]{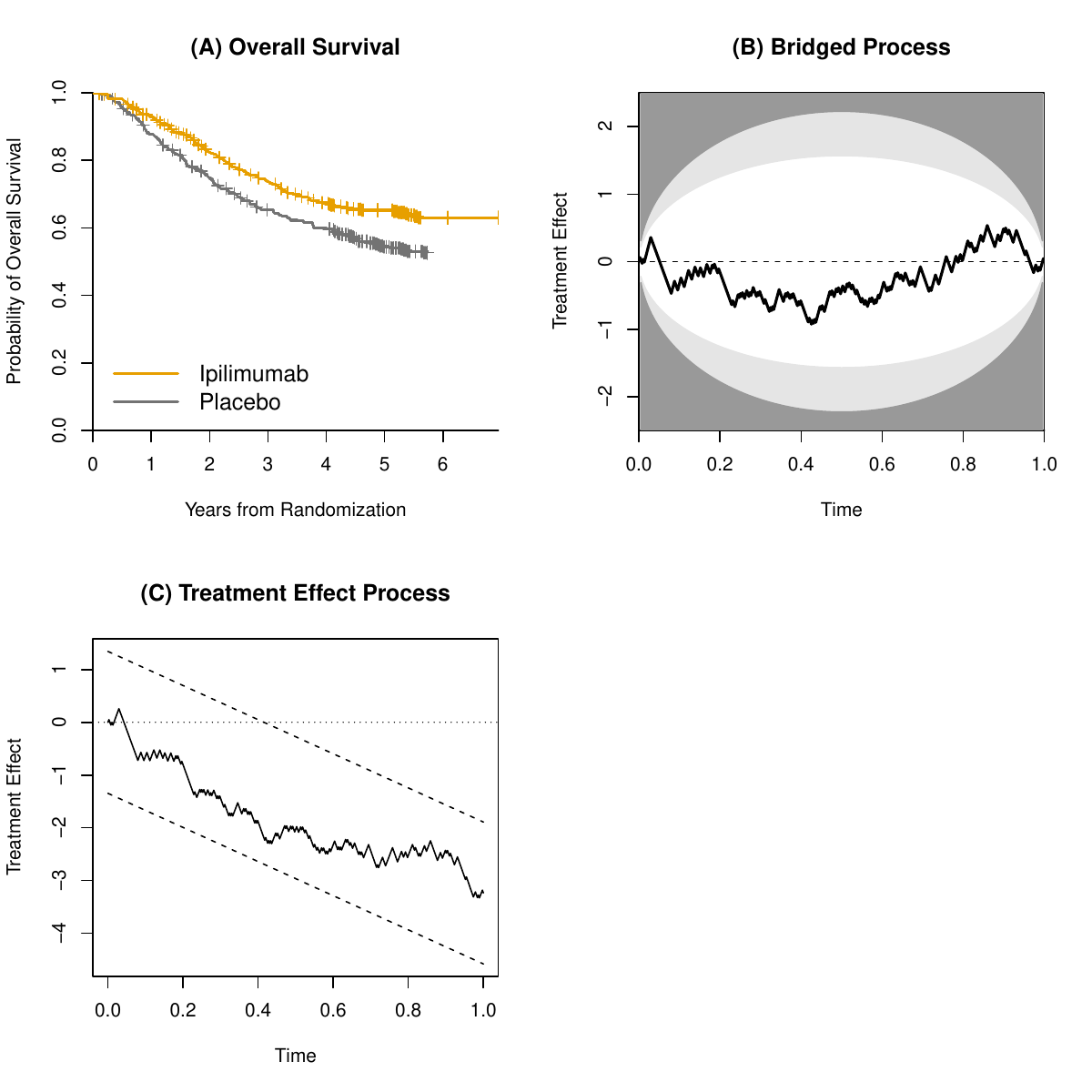}
 \caption{\label{fig.EggermontNEJM} (A) Estimated overall survival for patients treated with ipilimumab compared to placebo in \cite{Eggermont2016}. The bridge (B) and the treatment effect process (C) indicate no model inadequacy.}
\end{figure}

The overall survival curves are shown in Figure \ref{fig.EggermontNEJM} (A) for the digitized data set. The estimated hazard ratio for the treatment effect was 0.71 (95\% CI: 0.58 - 0.88). The bridged process in (B) and the regression effect process in (C) collectively indicate no notable deviation from proportional hazards. Therefore, an nph2ph transform is not required for this example. 

\section{Discussion}
Our approach in this work is very much a data analytic one. While our focus and development are built around the nph2ph transform, we limit our efforts to estimating this transform, which would, in practice, mean finding a good estimate of a time-transformed regression function, $\beta (t).$ A lot of work on finding good estimates of $\beta (t)$ has already been done \citep{Murphy1991, Murphy1993} and, rather than build on that, we set a more modest goal. This goal is not to identify as best we can the transform itself but to find a ready working approximation to it that we consider to be good enough in practice. For this to work, we need to define what we mean by ``good enough'' and we have taken it to be one that results in a good fit for the model. Although there are many formal tests of fit available and new tests could be built out of the bridged transform of the treatment effect process, we believe that our goal is reasonably well accomplished when this bridged process does not stray outside the white zone of the figures. Our conclusion on fit would be based on the ``eyeball factor'', does it look good, i.e., does it behave as we would wish so that the bridged process does not find itself in any shaded area. 

The examples and many others that we have studied tend to suggest that for situations in which the regression function $\beta (t)$ is either constant or slowly changing and not too volatile, a simple changepoint model will do enough to produce an improvement in fit that is good enough. This is backed up by the theoretical findings in \cite{jmoq03}, \cite{Chauvel2014}, and \cite{Chauvel2017} indicating that once we have achieved a good fit, any further improvements in fit will not correspond to noticeable changes in predictability. Let's not forget that all of this is conditional upon the regressors in the model. Moving away from a single binary covariate to more covariates may, of course, greatly impact predictability. That is another issue and not one we have looked into here. 

In a practical setting, the figures shown in the examples can add much to the discussion. Poor fit and the potential for significant improvements in predictability are immediately apparent. This is unlike many current discussions around non-proportionality that take the observed survival estimates, either the Kaplan-Meier or the Nelson-Aalen curves, as their starting point. When these curves overlap early on, only to diverge later, this will be taken to show delayed effects. When the curves begin to separate well only to later come back together, this will be taken to indicate a waning or a reversal of effects. We would not dispute that, but we do feel that the estimated survival curves are not the best visual tool to examine nonproportionality. This is very clearly shown in the \cite{LongNEJM2017}  data. In light of the observed Kaplan-Meier curves, most of us would take the proportional hazards assumption to be at least a reasonable approximation to the relation between the survival experience of the two groups. And yet, the treatment effect and the bridged process show the actual observations are poorly described by a proportional hazards assumption. 

\begin{figure}[h]
\centering
 \includegraphics[width=3in, height=3in]{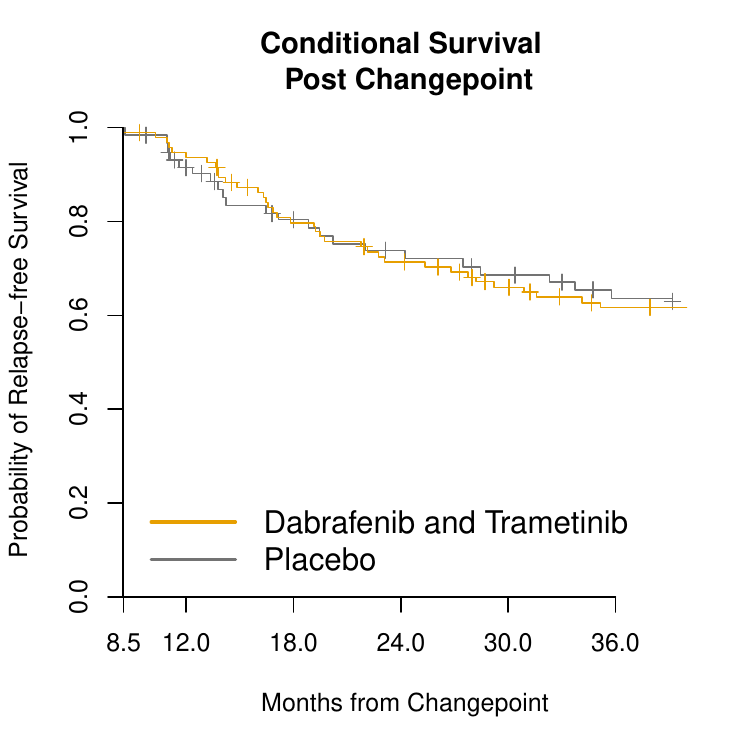}
 \caption{\label{fig.longLandmark} The estimated relapse-free survival landmarked at the changepoint of 8.5 months in the   \cite{LongNEJM2017} example. Conditioning on living to 8.5 months, there was no further indication of a treatment effect.}
\end{figure}

The consequences of violating the proportional hazards assumption in the \cite{LongNEJM2017} study are not minor. The overall conclusion that dabrafenib and trametinib are better than placebo is correct, but we only managed to arrive at this conclusion because the actual effects were so strong and the sample size was quite large. Had the effect been weaker - yet still very strong under a more accurate model specification - we may have missed the key scientific fact that dabrafenib and trametinib are better than placebo. This does not mean that, despite some shortcomings, we succeeded anyway. A key goal, detecting a treatment effect, was accomplished. But, the importance of this effect was largely underestimated. The treatment effect is, in fact, if only for the first 8.5 months, an order of magnitude greater than what we might have anticipated from working with the wrongly specified proportional hazards model. And this is only a part of the story. A nonproportional hazards model, much closer to the observations as indicated by the satisfactory fit (see Figure \ref{fig.longNEJM} D), suggests that, after week 8.5, there is almost no effect of treatment. This is easily confirmed by looking at the conditional survival estimates. Figure \ref{fig.longLandmark} shows that for all patients who survive until week 8.5 the remaining survival is almost independent of group status, the Kaplan-Meier curves all but overlap.  

Continuing our discussion on the \cite{LongNEJM2017} example, we might also note that from week 8.5 toward the study's closing date - a considerable percentage of the total study time - there was no treatment effect. Imagining that as time goes by and more events are recorded, we will increase the study's power, which can correspond to a misassumption in the design. If the model structure is ignored, our power can decrease with time rather than vice versa. In practice, the strong early effects, i.e., those observed before week 8.5, are becoming increasingly diluted. A mirror image of this situation can be seen in immunotherapy studies where an initial period may show either weak or no effect, followed by a steadily growing effect. Both situations mark significant departures from a proportional hazards assumption and must be addressed. How ought we respond to this in the design stage?  While we cannot know the future, little can be predicted with high precision, but we may be able to use similar or related studies analyzed retrospectively - specifically making use of the nph2ph transform. We can then consider design strategies that can help us increase the power, or at least not lose too much power, of the study that we are planning when nonproportional hazards is a real possibility.     

\section{Acknowledgements}
This work was supported by the NIH Grant P30CA008748.

\section{Data Availability}
The data included in the manuscript are based on digitized Kaplan-Meier survival curves from published clinical trials. The method detailed by Guyot et al (2012) was used to digitize each trial presented. 

\newpage
\clearpage

\bibliography{References.bib}

\end{document}